# The Basic Discrete Hilbert Transform with an Information Hiding Application

Renuka Kandregula

**Abstract:** This paper presents several experimental findings related to the basic discrete Hilbert transform. The errors in the use of a finite set of the transform values have been tabulated for the more commonly used functions. The error can be quite small and, for example, it is of the order of $10^{-17}$ for the chirp signal. The use of the discrete Hilbert transform in hiding information is presented.

## Introduction

The basic Discrete Hilbert Transform (DHT) of discrete data f(n) where n = (-∞,…,-1,0,1,…,∞) was given by Kak [1]:

$$DHT\{f(n)\} = g(k) = \begin{cases} \frac{2}{\pi}\sum_{n\ odd}\frac{f(n)}{k-n}; & k\ even \\ \frac{2}{\pi}\sum_{n\ even}\frac{f(n)}{k-n}; & k\ odd \end{cases} \qquad (1)$$

The inverse Discrete Hilbert Transform (DHT) is given as:

$$f(n) = \begin{cases} -\frac{2}{\pi}\sum_{k\ odd}\frac{g(k)}{n-k}; & n\ even \\ -\frac{2}{\pi}\sum_{k\ even}\frac{g(k)}{n-k}; & n\ odd \end{cases} \qquad (2)$$

The Hilbert transform has many applications in signal processing, imaging, modulation and demodulation, determination of instantaneous frequency and in cryptography [2],[3],[4],[5].

The discrete Hilbert transform (DHT) has several forms [6]-[9]. These forms use trigonometric functions and therefore they are relatively compute-intensive. For this reason we investigate the use of the basic DHT in this paper.

The specific application that we take here is that of information hiding to which DHT lends itself naturally since phase shift in speech makes no difference as far as perception is concerned [10]. For a background on analog information scrambling and hiding, see {11]-[14].

# Information Hiding Using DHT

In the method proposed in [10], the secret information, which could be an image or some other linear sequence, is coded in the binary form. The steganographic signal is hidden in the phase differences that will be produced based on whether the direct speech covert signal or its DHT has been transmitted.

Since DHT does not affect the spectrum, for it only shifts the phase and the human perception system is insensitive to it, the fact that the steganographic signal carries additional secret information will not be obvious.

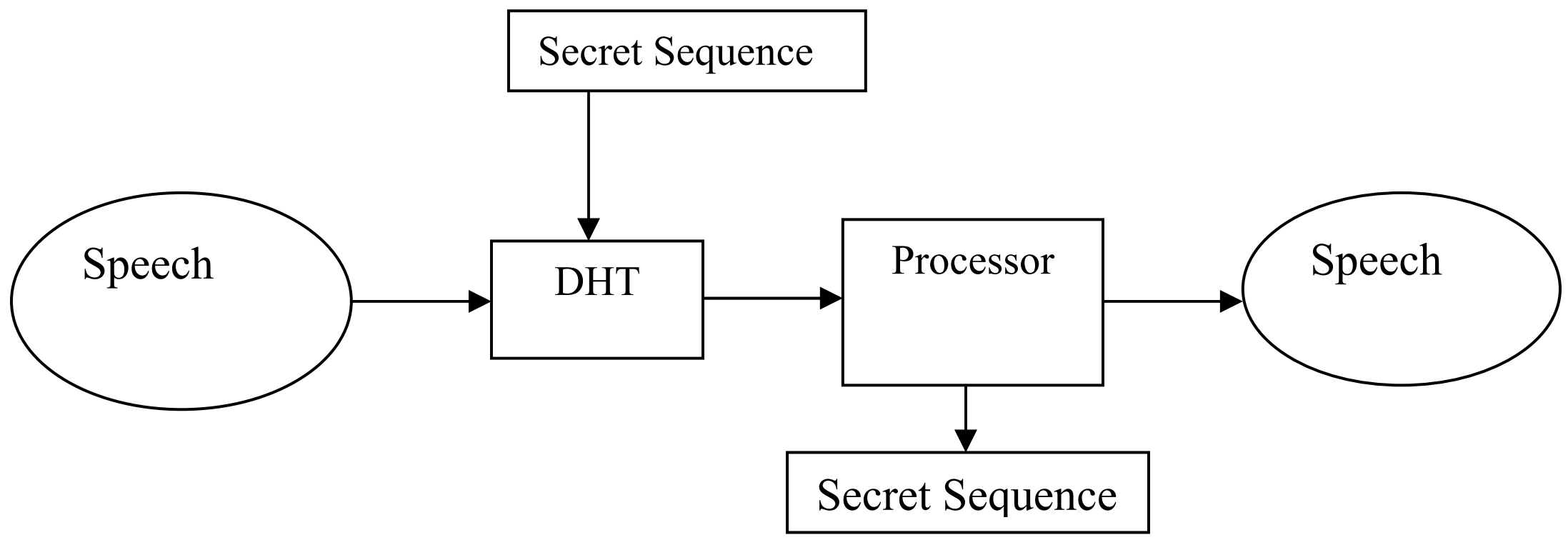

**Fig 1. Information Hiding System**

Figure 1 presents a schematic of the information hiding system. The secret sequence is determined by considering the DHT processed information for its shifts in phase according to a clock. The speech waveform will otherwise not be effected and therefore from a perception basis it will be unaltered.

# The Matrix Form of the DHT

The matrix form of the DHT requires that the data be of finite length. Since the DHT transform is defined for an infinite number of points, limitation of the DHT transform signal to a finite set would set up an approximation in the signal that is recovered.

The DHT is given below for data n=0, 1, 2, … :

$$
\begin{bmatrix} g(0) \\ g(1) \\ g(2) \\ g(3) \\ g(4) \\ g(5) \\ \cdot \\ \cdot \\ \cdot \end{bmatrix} = \frac{2}{\pi}
\begin{bmatrix}
0 & \frac{1}{-1} & 0 & \frac{1}{-3} & 0 & \frac{1}{-5} & 0 & \frac{1}{-7} & \cdot \\
\frac{1}{1} & 0 & \frac{1}{-1} & 0 & \frac{1}{-3} & 0 & \cdot & \cdot & \cdot \\
0 & \frac{1}{1} & 0 & \frac{1}{-1} & 0 & \frac{1}{-3} & \cdot & \cdot & \cdot \\
\frac{1}{3} & 0 & \frac{1}{1} & 0 & \frac{1}{-1} & 0 & \cdot & \cdot & \cdot \\
0 & \frac{1}{3} & 0 & \frac{1}{1} & 0 & \frac{1}{-1} & \cdot & \cdot & \cdot \\
\frac{1}{5} & 0 & \frac{1}{3} & 0 & \frac{1}{1} & 0 & \cdot & \cdot & \cdot \\
0 & \cdot & \cdot & \cdot & \cdot & \cdot & \cdot & \cdot & \cdot \\
\frac{1}{7} & \cdot & \cdot & \cdot & \cdot & \cdot & \cdot & \cdot & \cdot \\
\cdot & \cdot & \cdot & \cdot & \cdot & \cdot & \cdot & \cdot & \cdot
\end{bmatrix}
\begin{bmatrix} f(0) \\ f(1) \\ f(2) \\ f(3) \\ f(4) \\ f(5) \\ \cdot \\ \cdot \\ \cdot \end{bmatrix}
$$

The above can also be written as below:

$$
\begin{bmatrix} g(0) \\ g(1) \\ g(2) \\ g(3) \\ g(4) \\ g(5) \\ \cdot \\ \cdot \\ \cdot \end{bmatrix} = \frac{2}{\pi}
\begin{bmatrix}
0 & -1 & 0 & -3^{-1} & 0 & -5^{-1} & 0 & -7^{-1} & \cdot \\
1 & 0 & -1^{-1} & 0 & -3^{-1} & 0 & \cdot & \cdot & \cdot \\
0 & 1 & 0 & -1^{-1} & 0 & -3^{-1} & \cdot & \cdot & \cdot \\
3^{-1} & 0 & 1 & 0 & -1^{-1} & 0 & \cdot & \cdot & \cdot \\
0 & 3^{-1} & 0 & 1 & 0 & -1^{-1} & \cdot & \cdot & \cdot \\
5^{-1} & 0 & 3^{-1} & 0 & 1 & 0 & \cdot & \cdot & \cdot \\
0 & \cdot & \cdot & \cdot & \cdot & \cdot & \cdot & \cdot & \cdot \\
7^{-1} & \cdot & \cdot & \cdot & \cdot & \cdot & \cdot & \cdot & \cdot \\
\cdot & \cdot & \cdot & \cdot & \cdot & \cdot & \cdot & \cdot & \cdot
\end{bmatrix}
\begin{bmatrix} f(0) \\ f(1) \\ f(2) \\ f(3) \\ f(4) \\ f(5) \\ \cdot \\ \cdot \\ \cdot \end{bmatrix}
$$

We would like to show that the use of this form leads to very small error when the transformed DHT has the same size as the original size.

To verify the same, a number of inputs such as Sine, Cosine, Tangent, binary on and off with and without guard bands have been considered. The results obtained from the inverse transform on comparison with the original transform have a marginal Root Mean Square (RMS) error which shows that the inverse Discrete Hilbert Transform formula represented above holds good for any discrete data. The RMS error for ‘n’ input values is given by the formula:

$$\text{RMS error} = \sum_{n} \frac{(f(n) - F(n))^2}{n}$$

where f(n) = input values and F(n) = values obtained with inverse transform ($DHT^{-1}$ (DHT(f(n)) and n=total number of values considered.

To give a clear picture of what has been deduced, few examples have been shown in the following section of the paper.

# Results and Discussions

The following gives a detailed explanation and examples to support the statements that have been made in the earlier section of the paper. Upon implementing the MATLAB code, we have obtained the graphs that show that the original, Discrete Hilbert Transform of the Original and also the inverse of the Discrete Hilbert Transform.

The RMS error graphs and the average error computed show that the Inverse Discrete Hilbert Transform gives an output that differs marginally on comparison to the original input. The graphs shown below are obtained when the MATLAB code is run with and without a guard band. The guard band means that a range of values are pre-defined to be zero at the beginning and the end of the input. The purpose of this exercise is to determine the range of RMS errors that is obtained as against the regular means of plotting the graph in the case of without a guard band.

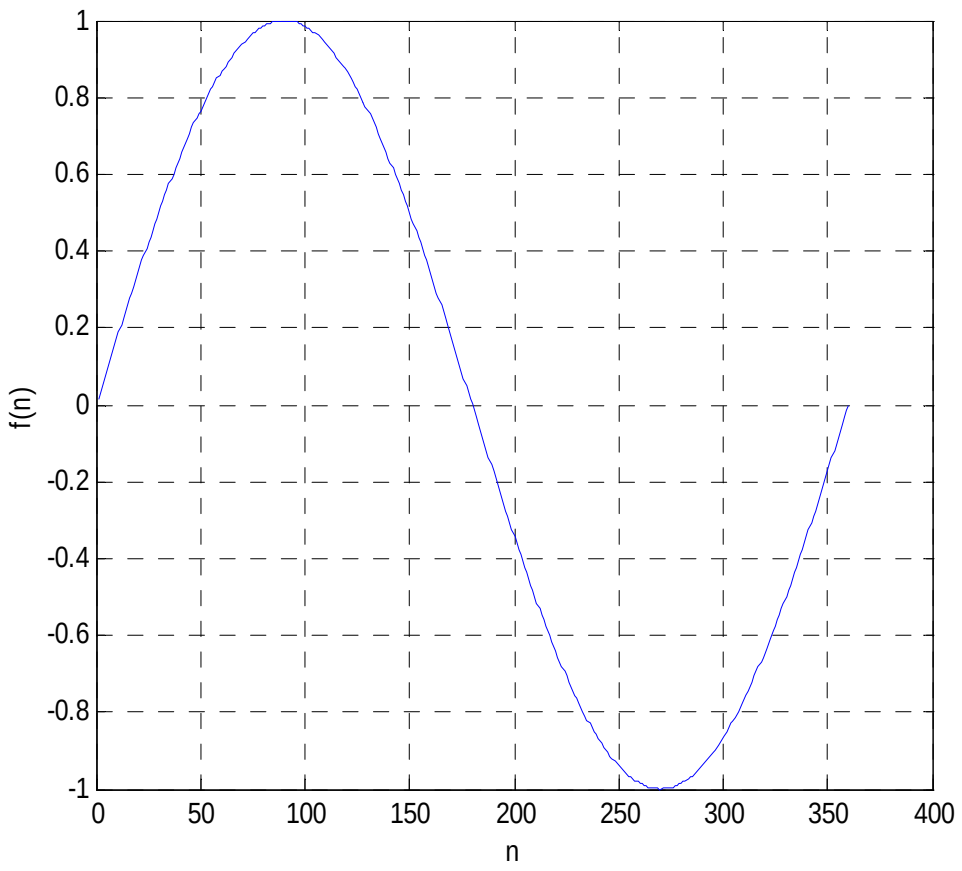

**Fig a**

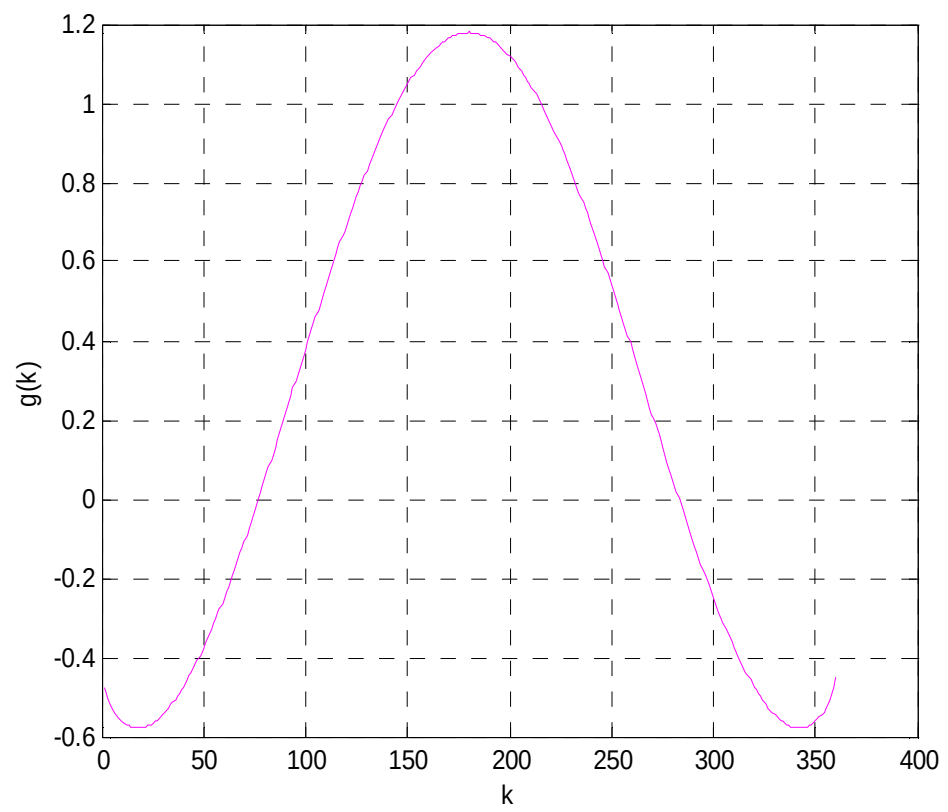

**Fig b**

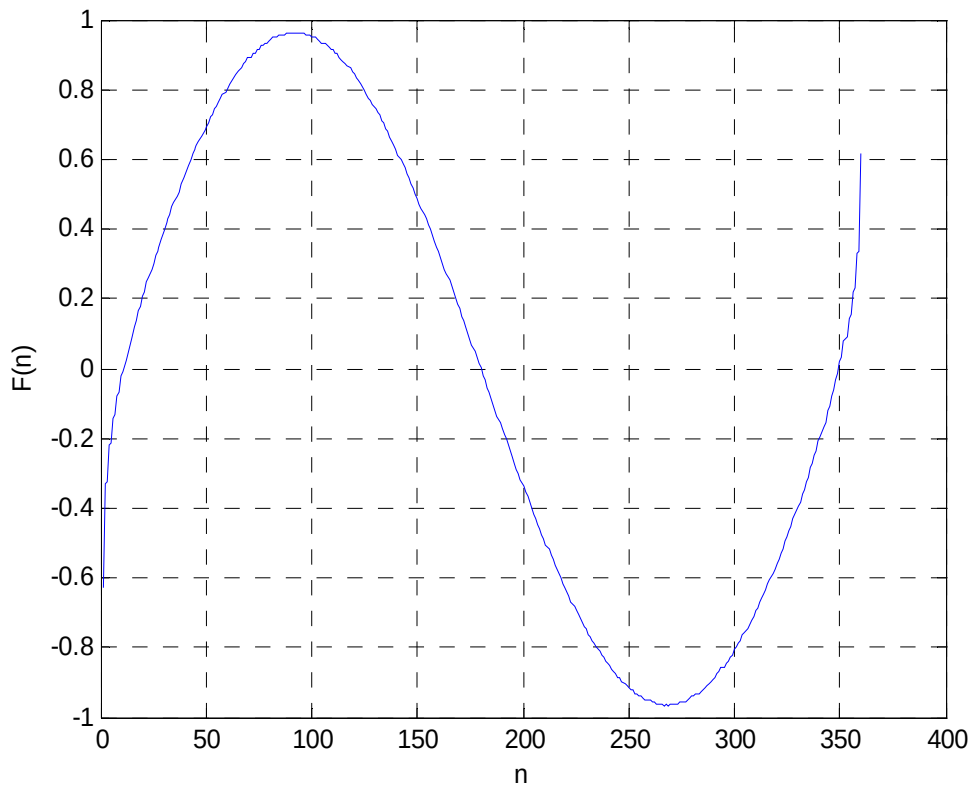

**Fig c**

Figure 1: Sine (a) Original, (b) DHT, (c) Inverse of the DHT

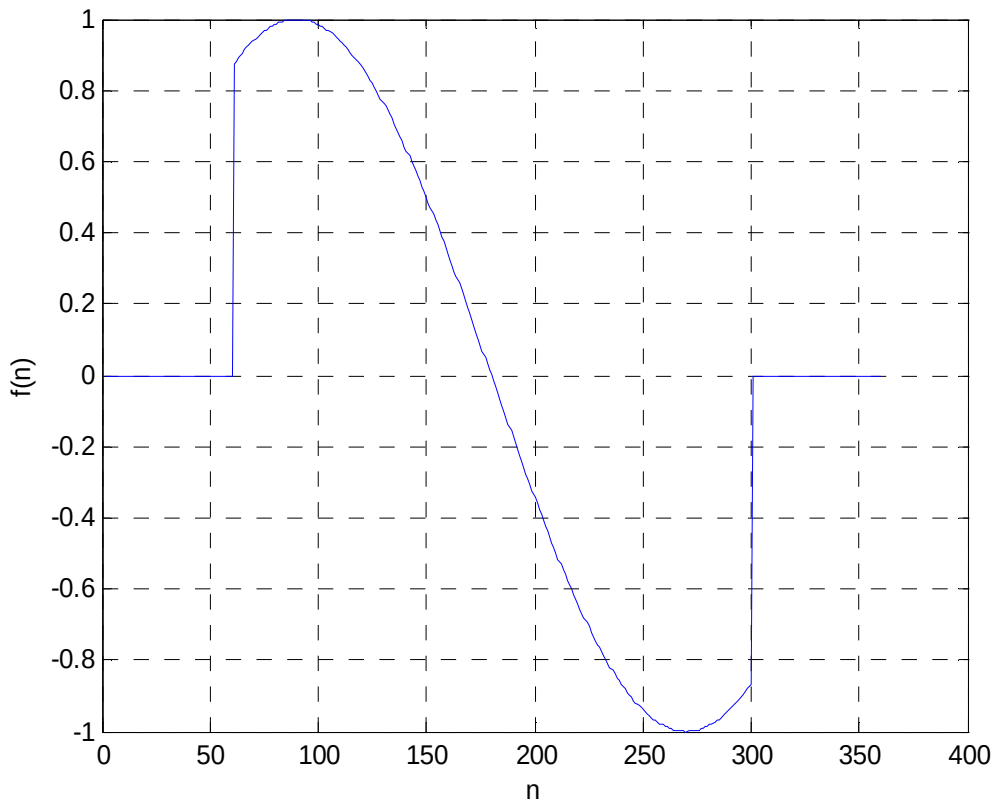

**Fig d**

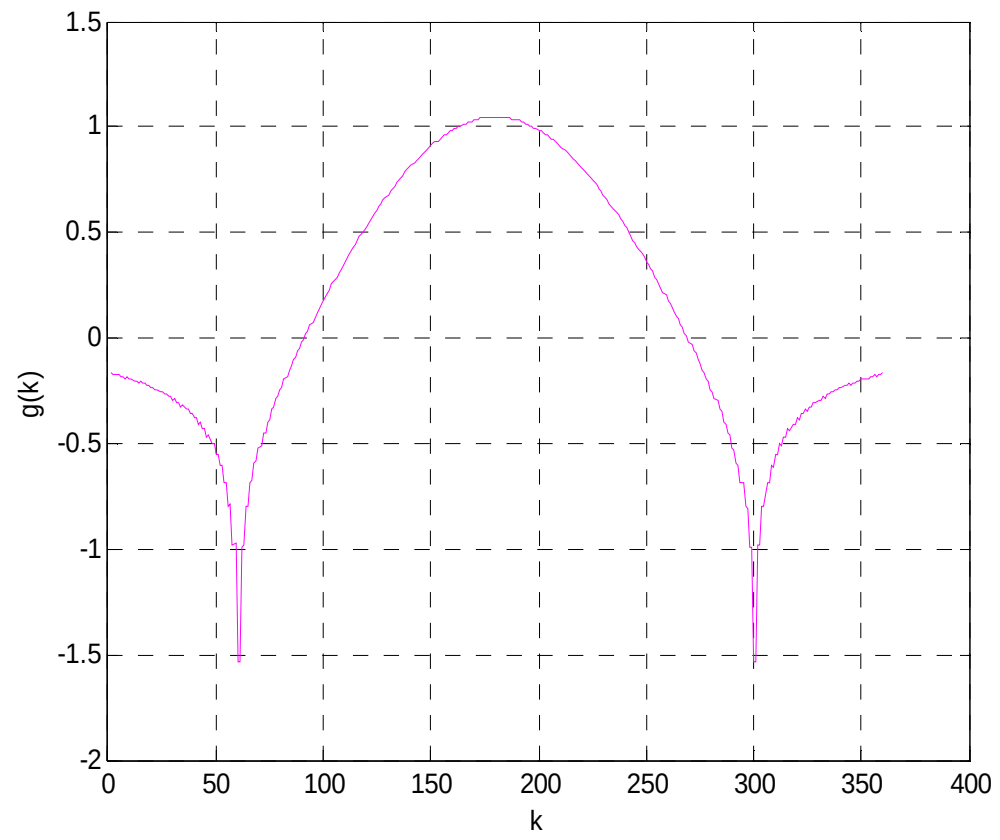

**Fig e**

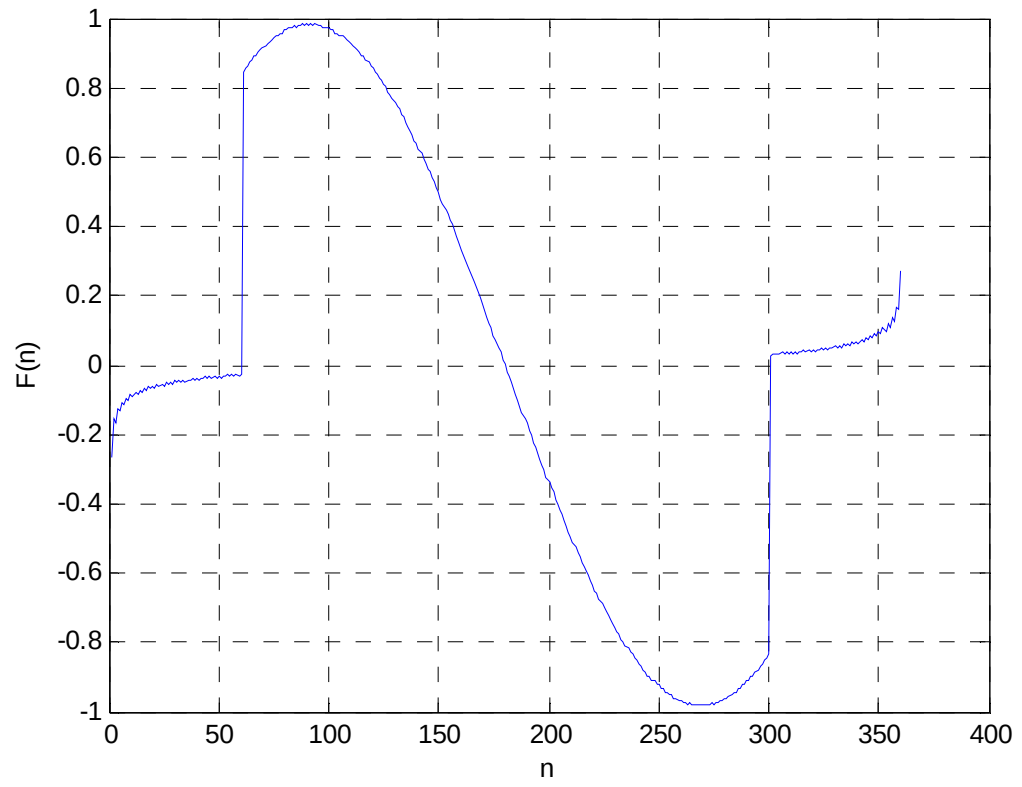

**Fig f**

Figure 2: Sine without Guard band (d) Original, (e) DHT, (f) Inverse of the DHT

This shows that the use of the guard band is not essential in the use of the DHT formulas since the error is very small.

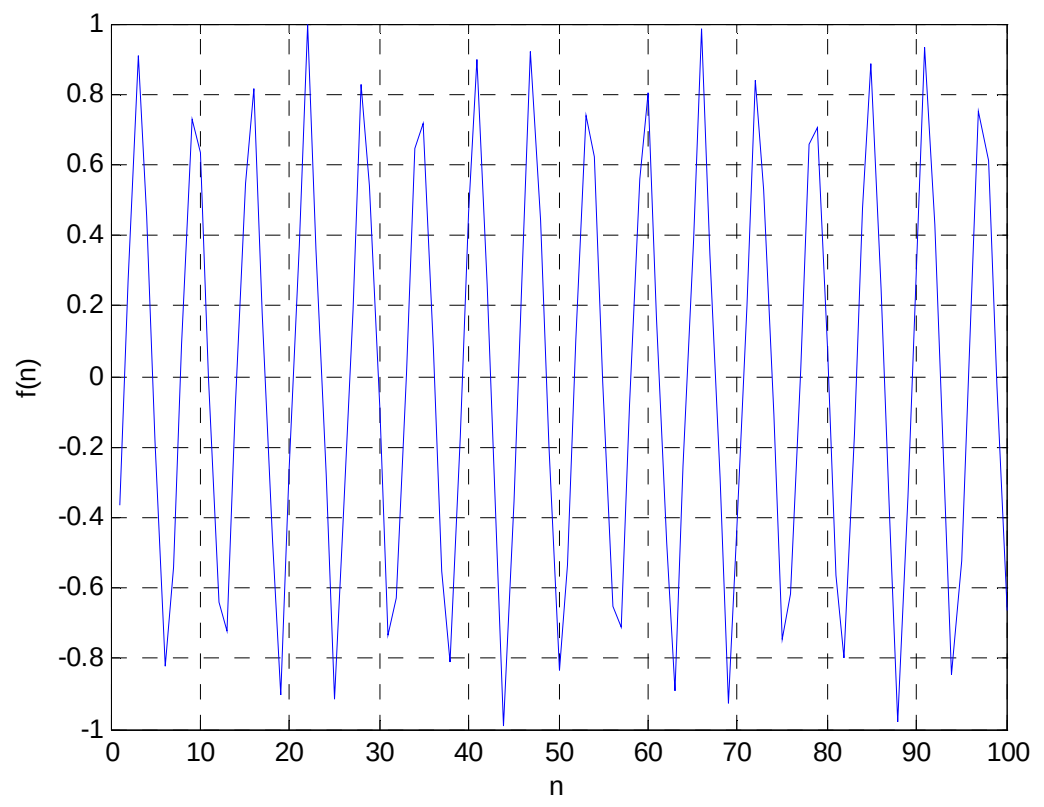

**Fig a**

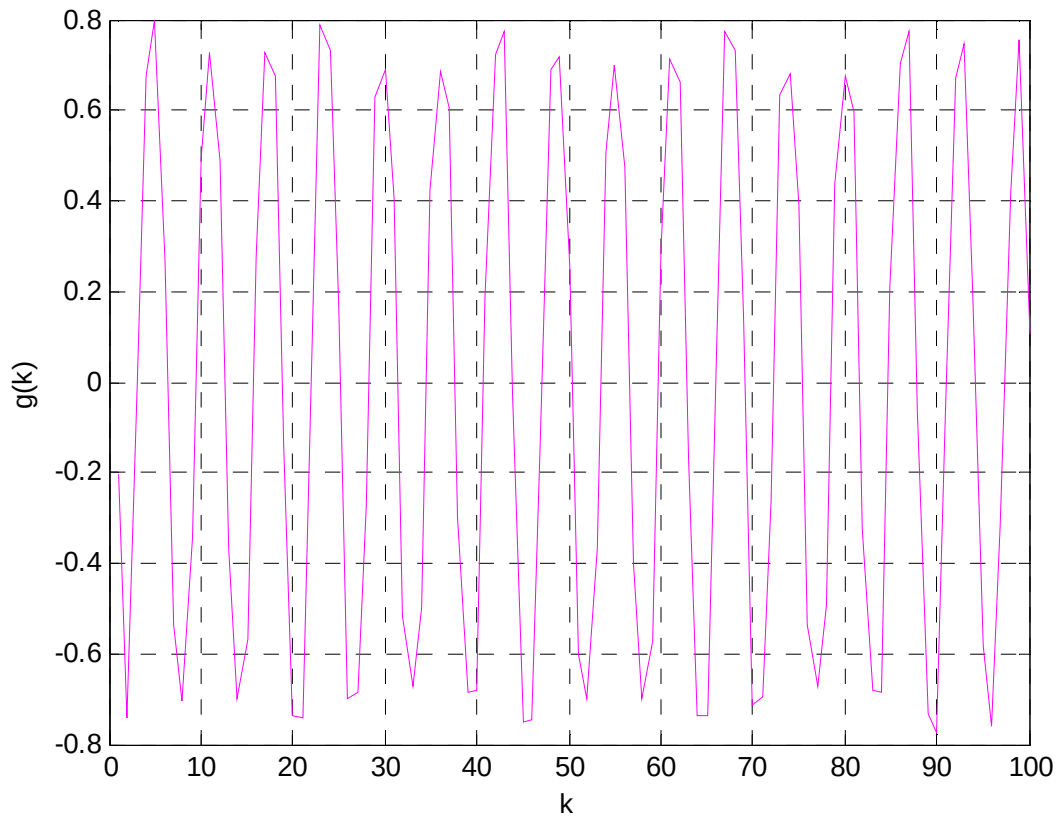

**Fig b**

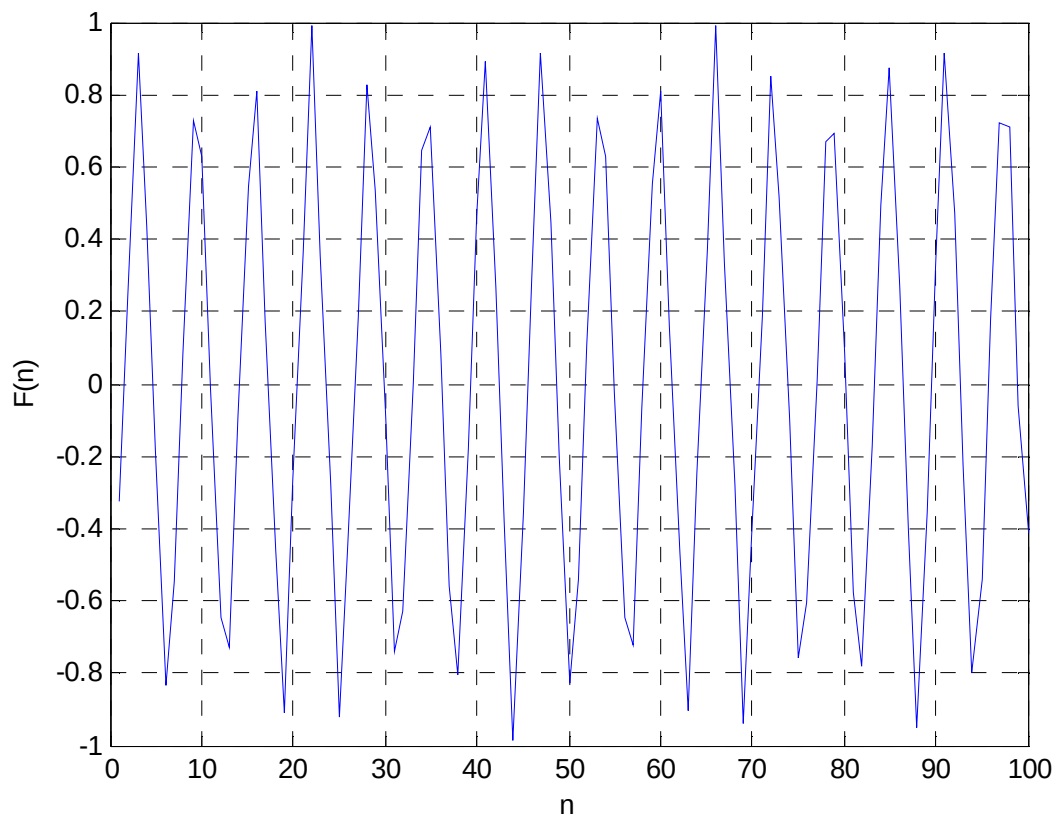

**Fig c**

Figure 3: Sawtooth (a) Original, (b) DHT, (c) Inverse of the DHT

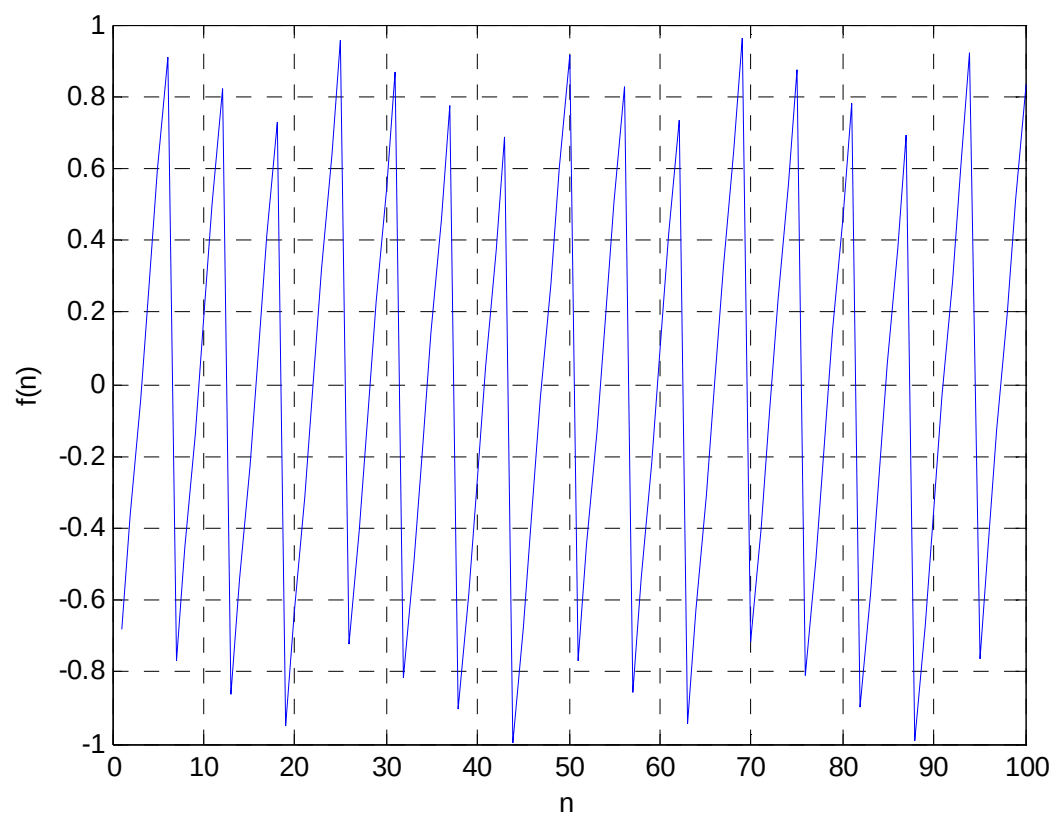

**Fig a**

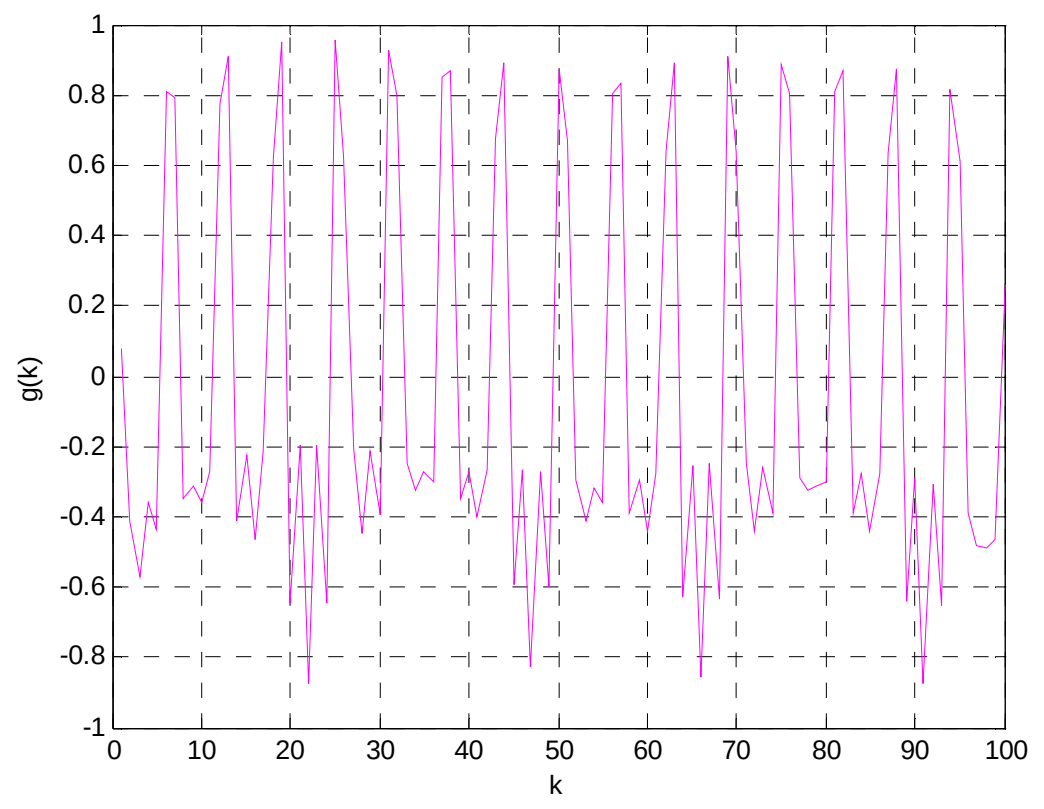

**Fig b**

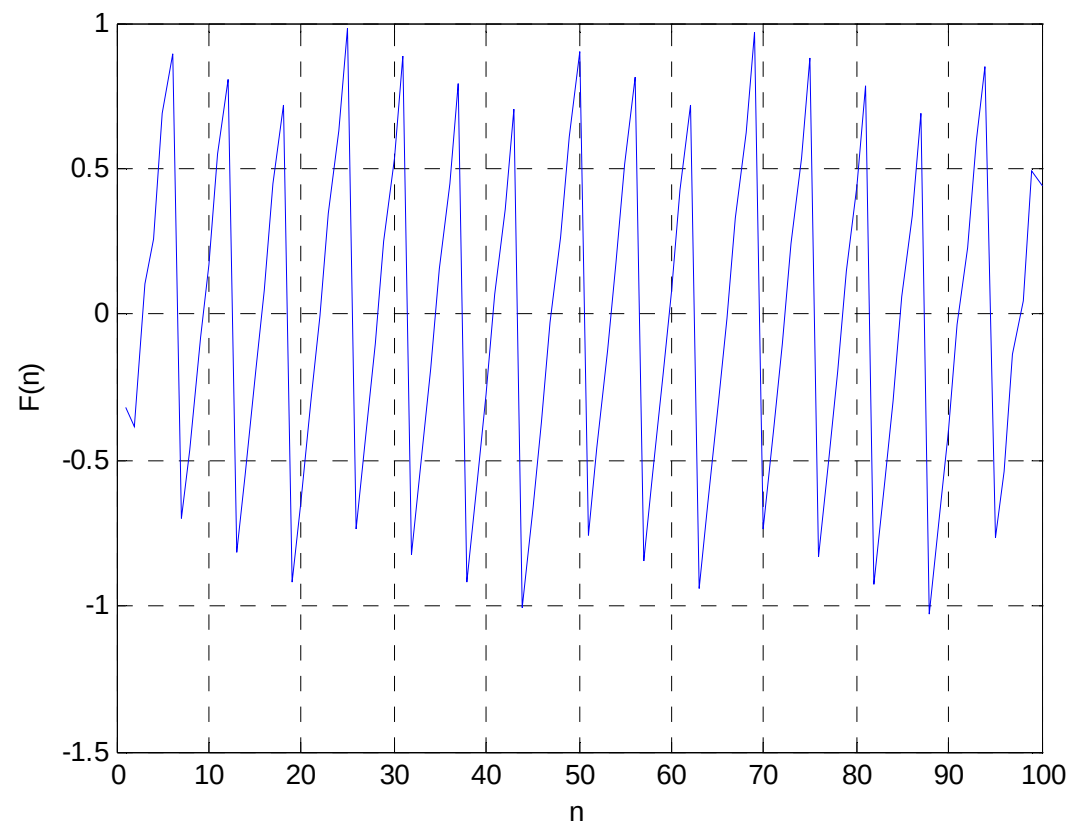

**Fig c**

Figure 4: Sawtooth (a) Original, (b) DHT, (c) Inverse of the DHT

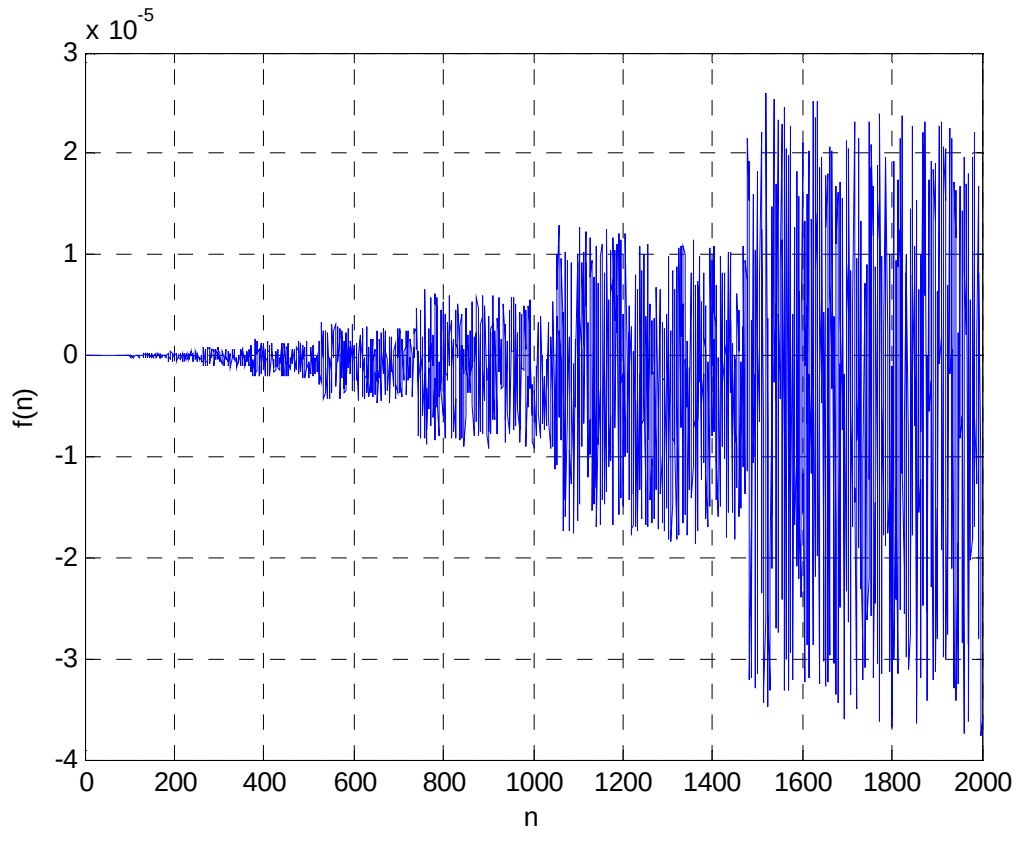

Fig g

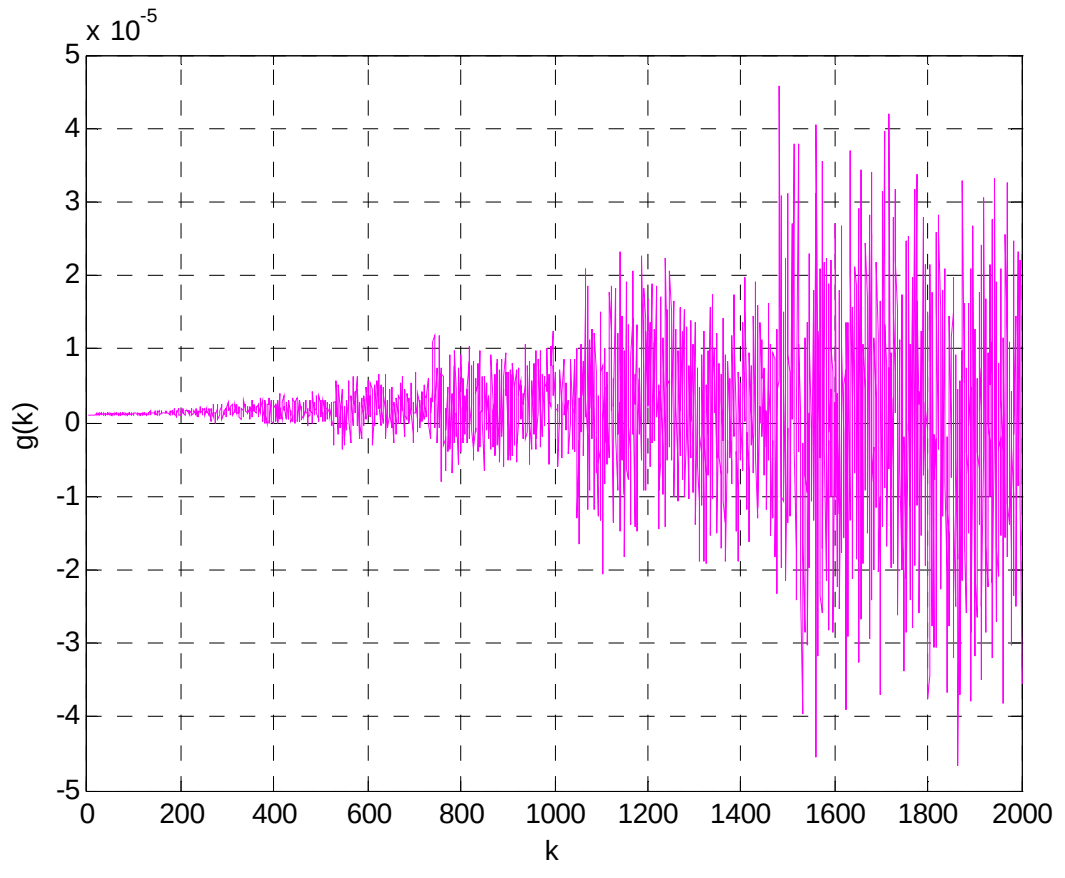

Fig h

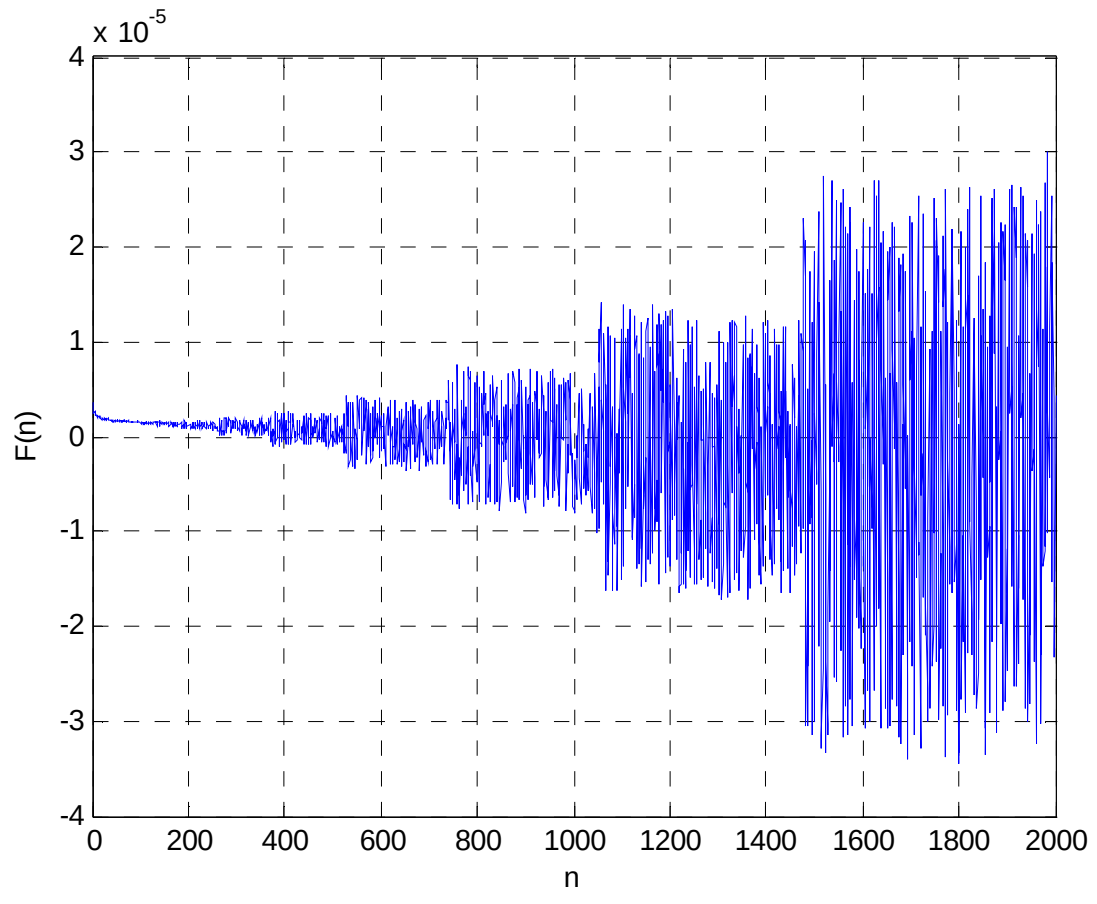

Fig i

Figure 5: Chirp (a) Original, (b) DHT, (c) Inverse of the DHT

We also present in Figure 6 the property of the DHT to subtract the average value of the signal which becomes an issue in images. In this example the DHT was computed on a line per line basis.

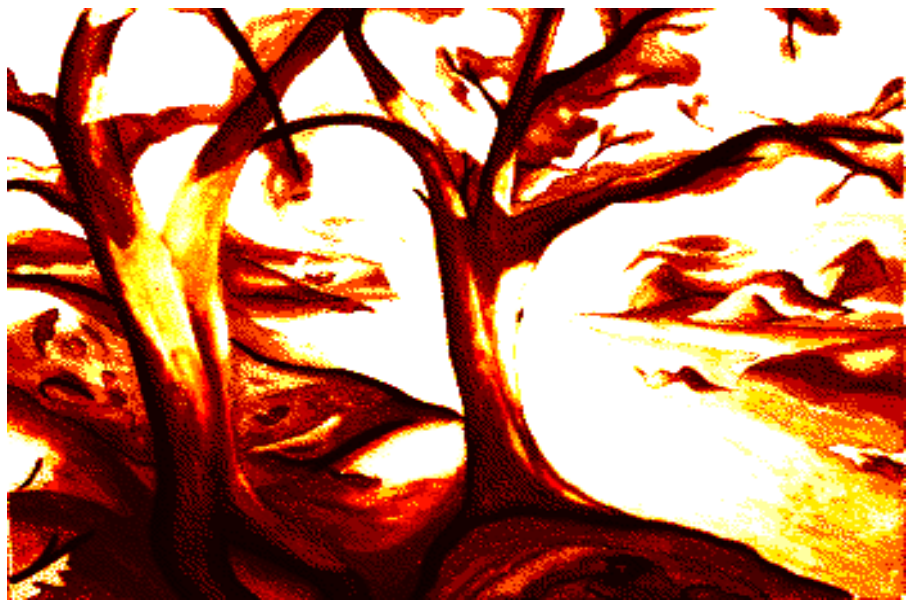

Fig a

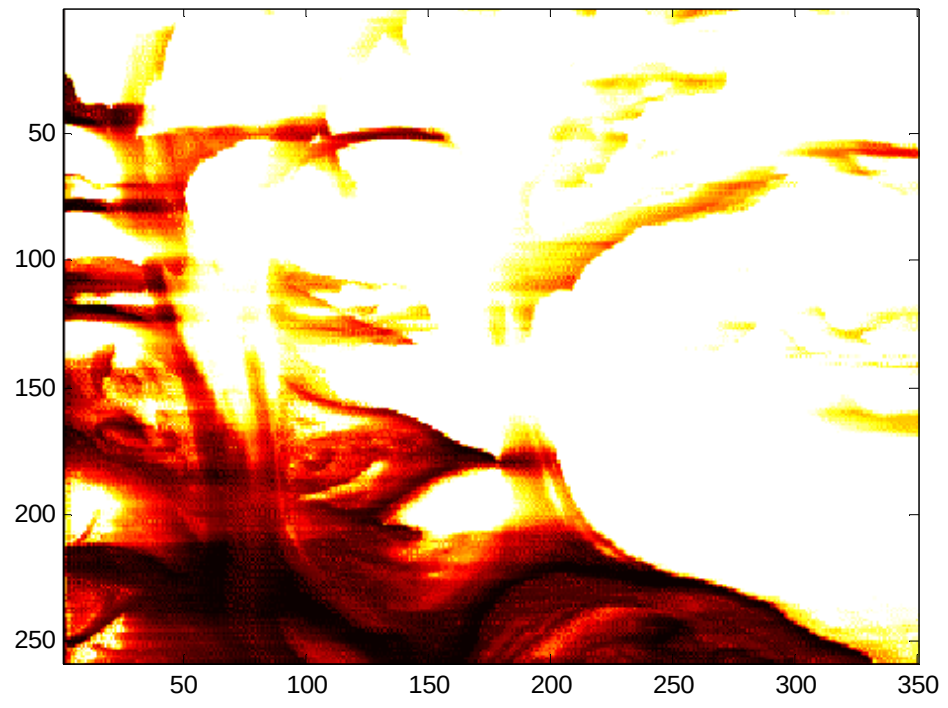

Fig b

Figure 6: Image (a) Original (b) DHT

## Average RMS Error over 'N' Values:

The average error is almost negligible which means that the Discrete Hilbert Transform and the Inverse Discrete Hilbert Transform holds good for any discrete data. Also we notice that the maximum error is at the beginning and at the end. An example of the Root Mean Square Error graph for a Chirp signal is shown below:

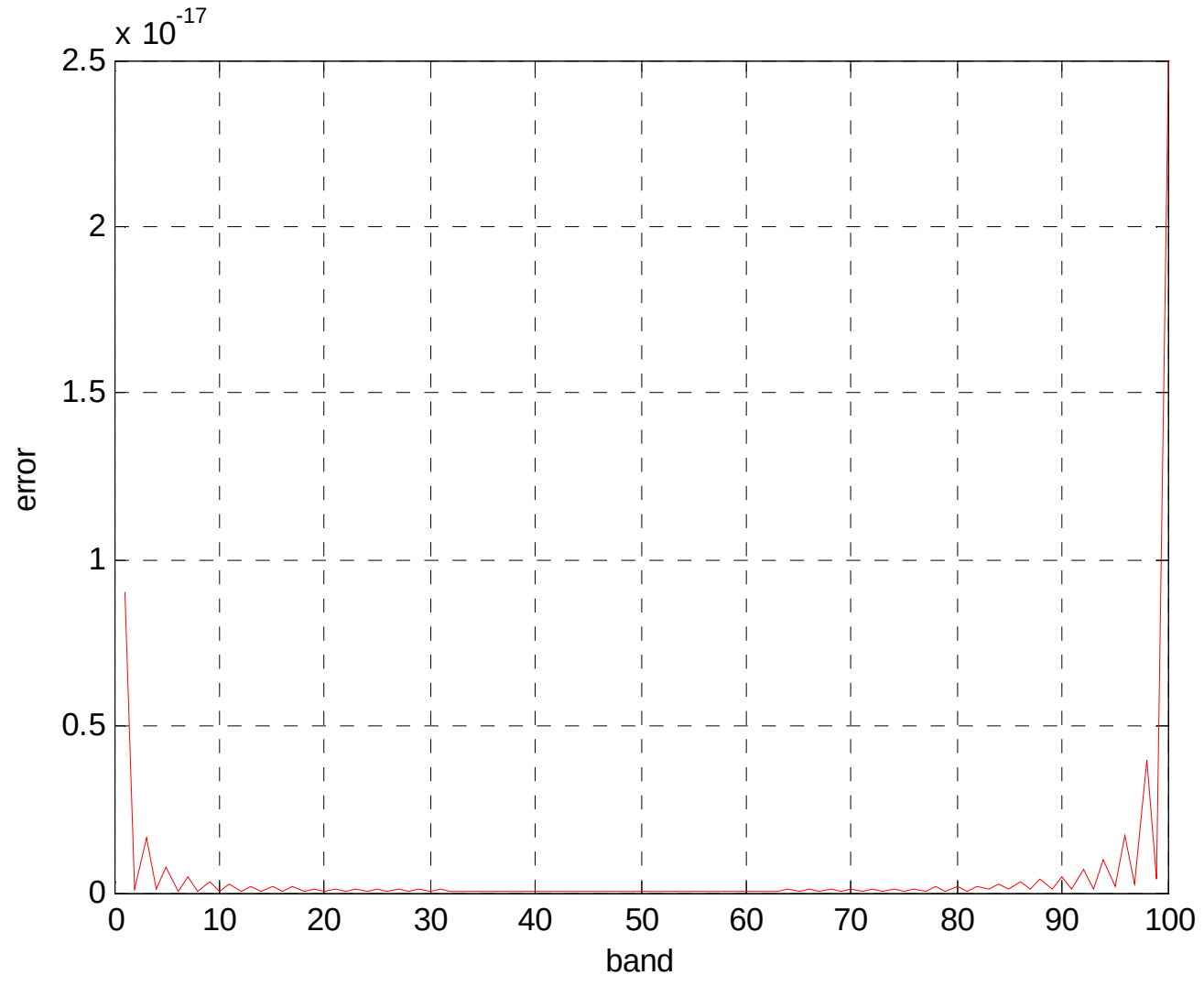

The average errors computed for different examples are as follows:

| S.No | Example | Average Root Mean Square Error |
|---|---|---|
| 1 | Sine | 2.9380e-004 |
| 2 | Sine with guard band | 5.7149e-005 |
| 3 | Cosine | 0.0018 |
| 4 | Cosine with guard band | 1.1651e-004 |
| 5 | Tangent with guard band | 0.0112 |
| 6 | On and Off | 0.0054 |
| 7 | Triangular | 2.0632e-033 |
| 8 | Sawtooth | 6.2457e-004 |
| 9 | Gauss Sinusoidal | 1.1081e-006 |
| 10 | Sawtooth with guard band | 1.1588e-007 |
| 11 | Dirichlet | 1.4372e-004 |
| 12 | Pulse Train | 2.4513e-004 |
| 13 | Pulse Train with guard band | 1.7924e-004 |
| 14 | Chirp | 2.4952e-017 |

## Conclusions

We have shown that the basis form of the DHT works very well for transformation of a variety of signals with error that is extremely small. In particular, the error for the chirp signal is of the order of $10^{-17}$.

This should increase the confidence of the use of this basic form in signal processing applications and also in applications of information hiding.

Issues that need further research include suitable forms to perform DHT for images.